\documentclass{article}
\PassOptionsToPackage{numbers}{natbib}
\usepackage{booktabs}
\usepackage{multirow}
\usepackage{amsmath}
\usepackage{enumitem}
\usepackage{graphicx}
\usepackage{float}

\usepackage[preprint]{neurips_2026}

\usepackage[utf8]{inputenc} 
\usepackage[T1]{fontenc}    
\usepackage{hyperref}       
\usepackage{url}            
\usepackage{booktabs}       
\usepackage{amsfonts}       
\usepackage{nicefrac}       
\usepackage{microtype}      
\usepackage{xcolor}         

\title{SemaVoice: Semantic-Aware Continuous Autoregressive Speech Synthesis}

\author{%
  \textbf{Huimeng Wang$^1$, Hui Lu$^1$, Jiajun Deng$^1$, Haoning Xu$^1$, Youjun Chen$^1$,} \\ \textbf{Xueyuan Chen$^1$, Zhaoqing Li$^1$, Shuhai Peng$^2$, Shiyin Kang$^{3\dagger}$, Xunying Liu$^1$} \\
  $^1$The Chinese University of Hong Kong, $^2$Tsinghua University, $^3$SenseTime Research\\
  \texttt{hmwang@link.cuhk.edu.hk} \\
}

\begin{document}

\maketitle

\begingroup
\renewcommand{\thefootnote}{\fnsymbol{footnote}}
\footnotetext[2]{Corresponding author.}
\endgroup

\begin{abstract}
Continuous autoregressive speech synthesis has recently emerged as a promising direction for zero-shot text-to-speech (TTS).
However, existing methods still suffer from a fundamental mismatch between semantic-prosodic modeling and reconstruction-driven continuous speech representations.
This mismatch causes TTS models to focus excessively on low-level acoustic textures at the expense of high-level semantic coherence, further exacerbating error accumulation in autoregressive generation.
To address this challenge, we propose SemaVoice, a semantic-aware continuous autoregressive framework for high-fidelity zero-shot TTS.
SemaVoice introduces a Speech Foundation Model (SFM) guided alignment mechanism that refines continuous speech representations to better capture both local semantic consistency and global structural relationships.
These representations condition a patch-wise diffusion head within the autoregressive framework for high-quality speech synthesis.
Experimental results on the Seed-TTS benchmark show that SemaVoice achieves an English WER of 1.71\% and remains highly competitive with state-of-the-art open-source systems in both objective and subjective evaluations.
The effectiveness of SFM guided alignment is further confirmed by significant improvements under varying representation granularities with a fixed information-rate constraint.
\end{abstract}




\section{Introduction}
Autoregressive (AR) language models (LMs) have recently emerged as a dominant paradigm in speech synthesis~\cite{borsos2023audiolm, kharitonov-etal-2023-speak, du2024cosyvoice, chen2025neural, wang2025spark, ye2025llasa, meng2025autoregressive}, demonstrating impressive generative capabilities and remarkable scaling properties.
Within this paradigm, text-to-speech (TTS) systems typically adopt discrete token prediction for high-fidelity zero-shot speech synthesis~\cite{du2024cosyvoice, chen2025neural, du2024cosyvoice2, du2025cosyvoice3, zhou2026indextts2}.
However, discrete tokens fundamentally impose a hard ceiling on reconstruction fidelity due to the irreversible loss of subtle acoustic details.
To mitigate this, a two-stage cascaded strategy is commonly employed: an LM first predicts coarse semantic tokens, after which a non-autoregressive (NAR) diffusion model restores fine acoustic details~\cite{du2024cosyvoice, du2024cosyvoice2, deng2025indextts}.
These cascaded models face a trade-off in simultaneously capturing long-range prosodic structures and fine-grained vocal richness, which limits the performance of speech synthesis.
Compared to discrete tokens, continuous representations reconstruct the original information more faithfully and provide a higher potential upper bound for synthesis quality~\cite{fanfluid, rombach2022high}.
Consequently, autoregressive modeling directly in a continuous speech representations space offers a more promising direction for high-fidelity speech synthesis.

Recent advancements have explored various frameworks for continuous autoregressive speech synthesis~\cite{meng2025autoregressive,lincontinuous,he2025continuous, wang2025felle, maefficient, ditar, vibevoice, zhou2026hierarchical}.
Pioneering works like Tacotron 2~\cite{shen2018natural} and recent models like MELLE~\cite{meng2025autoregressive} typically rely on standard regression objectives to generate mel-spectrograms.
Such regression-based approaches tend to favor mean predictions, often resulting in over-smoothed and low-diversity outputs~\cite{ren2022revisiting}.
To overcome this, continuous autoregressive TTS systems have increasingly incorporated diffusion denoising process~\cite{ho2020denoising, li2024autoregressive} to model the complex distributions of speech representations~\cite{he2025continuous, vibevoice, wu2025clear, ditar, zhou2026hierarchical}.
These systems exhibit remarkable perceptual quality and nuanced prosodic variations.
This expressive naturalness is driven by coupling high-level semantic-prosodic planning and fine-grained acoustic rendering within a unified modeling framework.
However, a fundamental mismatch persists between semantic-prosodic modeling and the reconstruction-driven continuous speech representations.
These representations are typically derived from Variational Autoencoders (VAEs)~\cite{Kingma2014} and thus inherently lack explicit alignment with text semantics.
This deficiency inevitably forces the TTS model to predominantly capture low-level acoustic textures at the expense of high-level semantic coherence.
Consequently, this compromise further exacerbates the widely observed error accumulation~\cite{pasini2024continuous}, which ultimately leads to degraded synthesis quality.

To tackle this challenge, we propose SemaVoice, a semantic-aware continuous autoregressive framework for high-fidelity text-to-speech synthesis.
SemaVoice aims to improve high-level semantic coherence while preserving rich acoustic textures in continuous autoregressive speech synthesis. 
The core of SemaVoice is a Speech Foundation Model (SFM)-guided alignment mechanism, which aligns continuous speech representations with a speech foundation model to enhance semantic consistency.
This mechanism explicitly addresses the mismatch between reconstruction-driven representations and semantic-prosodic modeling in continuous AR TTS.
In addition, we incorporate a high-compression $\sigma$-VAE representation~\cite{sun2024multimodal} and a patch-wise latent diffusion decoder with previous-latent conditioning to improve representation efficiency and generation stability.
This design reduces the autoregressive modeling burden by compressing sequence length and improving local generation stability. 
Collectively, SemaVoice provides a unified framework that achieves improved semantic-prosodic coherence and more stable autoregressive generation for high-fidelity text-to-speech synthesis.
Our main contributions are summarized as follows:
\begin{itemize}[leftmargin=2.5em, itemsep=4pt, topsep=2pt, parsep=0pt, partopsep=0pt]
    \item We present SemaVoice, a semantic-aware continuous autoregressive modeling framework that enables effective unified modeling of high-level semantic-prosodic planning and fine-grained acoustic rendering for high-fidelity zero-shot TTS.
    \item We identify the mismatch between reconstruction-driven continuous speech representations and semantic-prosodic modeling as a key bottleneck in continuous autoregressive TTS. To address this issue, we introduce an SFM-guided alignment mechanism that enhances semantic coherence by aligning continuous representations with high-level semantic and prosodic information encoded in speech foundation models, improving generation robustness without modifying the downstream TTS architecture or training pipeline.
    \item We validate SemaVoice through large-scale training on 150K hours of bilingual speech, achieving competitive performance compared with state-of-the-art open-source zero-shot TTS systems  in both objective and subjective evaluations.
\end{itemize}

\section{Related Work}
\subsection{Zero-shot Text-to-speech}
Existing zero-shot TTS systems can be broadly divided into autoregressive (AR) and non-autoregressive (NAR) paradigms.
Autoregressive approaches formulate speech generation as sequence modeling task.
In contrast, non-autoregressive models, such as Voicebox~\cite{le2023voicebox}, generate acoustic representations in parallel, achieving efficient inference while maintaining high fidelity.
Diffusion models have become a key component in NAR TTS, as demonstrated by E2-TTS~\cite{eskimez2024e2}, E3-TTS~\cite{gao2023e3}, F5-TTS~\cite{chen2025f5}, and NaturalSpeech~\cite{tan2024naturalspeech,shen2023naturalspeech, ju2024naturalspeech}.
Recent works like~\cite{niu2025semantic, cheng2026distillation} further introduce semantic alignment regularization to mitigate the trade-off between reconstruction fidelity and generation intelligibility in NAR TTS.
In contrast, our work focuses on autoregressive continuous speech modeling, where alignment stabilizes semantic planning under sequential generation.

\subsection{Discrete Autoregressive Text-to-speech}
Discrete autoregressive language models exhibit strong zero-shot and in-context learning capabilities in speech synthesis. 
VALL-E~\cite{chen2025neural} formulates TTS as conditional language modeling over discrete speech tokens, inspiring a series of neural codec-based TTS systems~\cite{han2024vall, nishimura2024hall, song2025ella, xin2024rall}. 
However, the limited bitrate of discrete tokens restricts fine-grained acoustic reconstruction. 
To mitigate this, cascaded approaches such as CosyVoice~\cite{du2024cosyvoice, du2024cosyvoice2, du2025cosyvoice3}, BASE-TTS~\cite{lajszczak2024base}, FireRed TTS~\cite{guo2024fireredtts}, and Seed-TTS~\cite{anastassiou2024seed} first predict coarse semantic tokens and then refine them using flow or diffusion models. 
Nevertheless, these cascaded pipelines still struggle to jointly model long-range prosody and fine acoustic detail.

\subsection{Continuous Autoregressive Text-to-speech}
Continuous autoregressive (AR) TTS models offer an alternative approach for high-fidelity speech synthesis.
MELLE~\cite{meng2025autoregressive} predicts Mel-spectrograms directly from text, and further incorporates variational inference to facilitate sampling mechanisms.
Recent continuous AR models increasingly integrate diffusion-based denoising processes to enhance generation diversity~\cite{vibevoice, ditar, zhou2026hierarchical}.
These approaches unify semantic-prosodic planning with local acoustic rendering, resulting in nuanced prosody and high perceptual quality.
DiTAR~\cite{ditar} extends this line of work with a patch-wise generation strategy, while VibeVoice~\cite{vibevoice} incorporates an acoustic-semantic hybrid tokenizer into a next-token diffusion framework for podcast generation.
Later studies, such as FELLE~\cite{wang2025felle} and VoxCPM~\cite{zhou2026hierarchical}, explore coarse-to-fine hierarchical modeling in continuous AR TTS.
Next-distribution prediction~\cite{xia2026kall} has also been successfully applied for improved efficiency and quality.
Despite these advances, the intrinsic mismatch between semantic-prosodic planning and reconstruction-driven speech representations still constrains the fidelity of generated speech.

\section{Method}
\subsection{SFM-Guided Alignment for Continuous Speech Representations}
\label{sec:sigma_vae}

\subsubsection{Continuous Speech VAE Formulation}
SemaVoice is built upon a continuous autoregressive speech synthesis framework over speech representations learned by a variational autoencoder (VAE)~\cite{Kingma2014}.
In this framework, the quality of the representation space is crucial for effective autoregressive generation.
Inspired by~\cite{sun2024multimodal}, we adopt the $\sigma$-VAE variant to achieve high speech compression while maintaining high-fidelity speech reconstruction.

\textbf{Architecture and Formulation}:
As shown in Fig.~\ref{fig:framework} (a), our speech VAE comprises an encoder $\mathcal{E}$ and a decoder $\mathcal{D}$.
Given a 24kHz input speech sequence $\mathbf{x}$, the encoder maps it to the mean parameter $\boldsymbol{\mu}$ of the latent distribution.
Distinct from vanilla VAEs, the $\sigma$-VAE introduces a stochastic variance parameter $\boldsymbol{\sigma}$, sampled from a pre-defined distribution $\mathcal{N}(0, C_\sigma)$ (see Eq.~(1)).
This formulation ensures that the representation space maintains a consistent and non-vanishing variance, providing a more stable representation for downstream autoregressive modeling~\cite{sun2024multimodal}.
The latent representation $\mathbf{z}$ is sampled via the reparameterization trick:
\begin{align}
    \mathbf{z} &= \boldsymbol{\mu} + \boldsymbol{\sigma} \odot \boldsymbol{\epsilon}, \quad \text{where } \boldsymbol{\epsilon} \sim \mathcal{N}(0, \mathbf{I}), \boldsymbol{\sigma} \sim \mathcal{N}(0, C_\sigma)
\end{align}
where $\odot$ denotes the element-wise product. Then the decoder $\mathcal{D}$ reconstructs the waveform $\hat{\mathbf{x}}$ from $\mathbf{z}$.
Following~\cite{kumar2023dac}, a multi-task learning objective with a discriminator setting is employed and can be expressed as
\begin{align}
    \mathcal{L}_{VAE} = \lambda_{mel}\cdot\mathcal{L}_{mel} + \lambda_{fm}\cdot\mathcal{L}_{fm} + \lambda_{adv}\cdot\mathcal{L}_{adv}+\lambda_{kl}\cdot\mathcal{L}_{kl}
\end{align}
where $\mathcal{L}_{mel}$ is the multi-resolution Mel-spectrogram reconstruction loss. 
$\mathcal{L}_{adv}$ and $\mathcal{L}_{fm}$ denote adversarial and feature matching losses from the discriminator, respectively, which improve sample fidelity and training stability. 
$\mathcal{L}_{kl}$ denotes the KL divergence that regularizes the latent space towards the prior distribution.
We empirically set $\lambda_{mel}=15.0$, $\lambda_{fm}=2.0$, $\lambda_{adv}=1.0$, and $\lambda_{kl}=0.01$.

\subsubsection{SFM Guided Alignment}\label{sec:sfm_alignment}
\begin{figure*}[t]
    \centering
    \includegraphics[width=\textwidth]{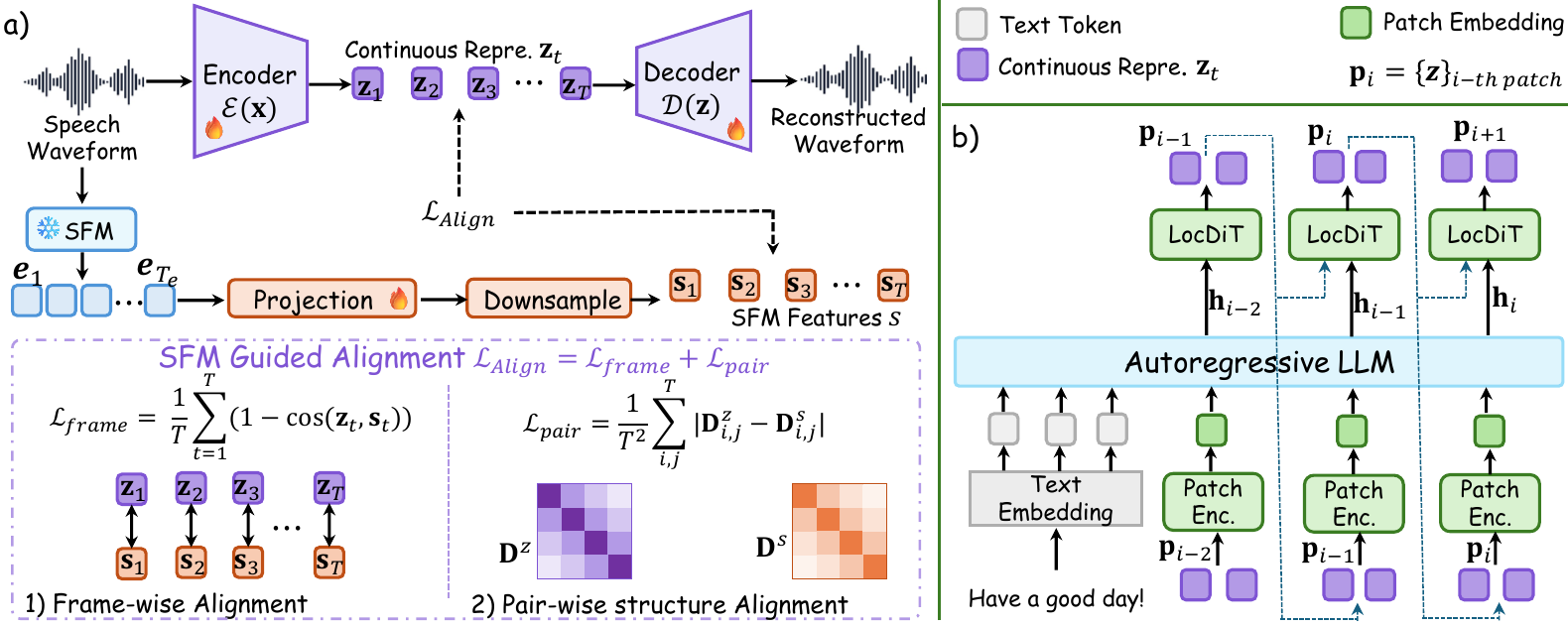}
    \caption{
    Overview of the proposed SemaVoice framework.
    (a) Speech Foundation model (SFM) guided alignment in VAE training. We illustrate only the computation of the SFM-guided alignment loss, while other standard VAE training losses are omitted for clarity. $\mathbf{D}^z$ and $\mathbf{D}^s$ denote self-similarity matrices.
    (b) Continuous autoregressive inference with next-patch diffusion for speech synthesis.
    }
    \label{fig:framework}
\end{figure*}

Continuous autoregressive speech synthesis requires jointly modeling high-level semantic-prosodic planning and fine-grained acoustic rendering.
However, speech representations learned by VAEs are primarily optimized for reconstruction fidelity, and therefore often under-represent the semantic information required for downstream semantic-prosodic modeling.
To address this mismatch, we introduce a speech foundation model (SFM)-guided semantic alignment mechanism in SemaVoice for high-fidelity speech synthesis.

As illustrated in Fig.~\ref{fig:framework}(a), given an input speech sequence $\mathbf{x}$, we extract semantic features $\mathbf{e}=\mathcal{F}(\mathbf{x})$ from a frozen SFM encoder $\mathcal{F}$, where $\mathbf{e}\in\mathbb{R}^{T_e \times d_e}$ provides high-level semantic information learned from large-scale self-supervised pretraining.
In parallel, the VAE encoder produces continuous representations $\mathbf{z}$ for speech reconstruction.
Since the SFM features $\mathbf{e}$ and speech representations $\mathbf{z}$ differ in both feature dimension and temporal resolution, we first project $\mathbf{e}$ into the same feature space as $\mathbf{z}$ and then align their temporal resolutions via linear interpolation.
Specifically, the transformed SFM features are denoted as $\mathbf{s}=\mathcal{P}_e(\mathbf{e})$, where $\mathbf{s}\in\mathbb{R}^{T\times d}$ has the same sequence length and feature dimension as the VAE latent representations $\mathbf{z}\in\mathbb{R}^{T\times d}$.

Following prior work~\cite{yao2025reconstruction}, we adopt a marginal consistency loss as the alignment objective in the proposed semantic alignment mechanism.
Specifically, the alignment loss is defined as 
\begin{align}
\mathcal{L}_{align}=\mathcal{L}_{frame}+\mathcal{L}_{pair}
\end{align}
where the two terms correspond to frame-wise alignment for local semantic consistency and pair-wise structure alignment for preserving global temporal relationships encoded by the SFM features.

\textbf{Frame-wise Alignment}:
We first enforce local semantic consistency by penalizing insufficient frame-wise similarity between the VAE speech representations $\mathbf{z}$ and transformed SFM features $\mathbf{s}$:
\begin{align}
\mathcal{L}_{frame}
=
\frac{1}{T}\sum_{t=1}^{T}
\left(
1-\cos(\mathbf{z}_t,\mathbf{s}_t)
\right)
\end{align}
where $\cos(\mathbf{z}_t,\mathbf{s}_t)$ denotes the cosine similarity between $\mathbf{z}_t$ and $\mathbf{s}_t$ at time step $t$. This term encourages each latent frame to preserve semantically meaningful local information.

\textbf{Pair-wise Structure Alignment}:
To further preserve global structural relationships across time steps, we align the self-similarity structures of the VAE representations $\mathbf{z}$ and semantic sequences $\mathbf{s}$.
We first compute the pair-wise cosine similarity matrices $\mathbf{D}^{z}, \mathbf{D}^{s}\in\mathbb{R}^{T\times T}$, where $\mathbf{D}^{z}_{i,j}=\cos(\mathbf{z}_i,\mathbf{z}_j)$ and $\mathbf{D}^{s}_{i,j}=\cos(\mathbf{s}_i,\mathbf{s}_j)$ measure the similarity between two time steps within the latent and semantic sequences, respectively. The pair-wise structure alignment loss is then defined as
\begin{align}
\mathcal{L}_{pair}
=
\frac{1}{T^2}
\sum_{i,j}
\left|\mathbf{D}^{z}_{i,j}-\mathbf{D}^{s}_{i,j}\right|
\end{align}
This term encourages preserving the global relational structure encoded by the SFM features.

\textbf{Adaptive Weighting}:
We employ an adaptive weighting strategy for the SFM-guided alignment loss based on gradient magnitude ratios computed on a shared encoder parameter set $\theta$:
\begin{align}
\lambda_{align}
=
\alpha \cdot
\frac{\left\|\nabla_{\theta}\mathcal{L}_{mel}\right\|_2}
{\left\|\nabla_{\theta}\mathcal{L}_{align}\right\|_2+\epsilon}
\end{align}
where $\alpha$ is a base scaling coefficient set to $0.5$, and $\epsilon$ is a small constant for numerical stability.

After incorporating SFM guided alignment into VAE formulation,  the complete training objective is:
\begin{align}
\mathcal{L}_{total}
=
\mathcal{L}_{VAE}
+
\lambda_{align}\mathcal{L}_{align}.
\end{align}

\subsection{Continuous Autoregressive Modeling for Speech Synthesis}
\label{sec:diffusion_head}
\subsubsection{Autoregressive Modeling with LLM and Diffusion Head}
As shown in Fig.~\ref{fig:framework} (b), SemaVoice follows the next-token diffusion formulation~\cite{li2024autoregressive} for autoregressive speech synthesis from continuous representations: an LLM backbone for sequence modeling and a diffusion head for continuous representation generation.

\textbf{Continuous Autoregressive Formulation}:
Let $\mathbf{z} = (\mathbf{z}_1, \mathbf{z}_2, \dots, \mathbf{z}_T)$ denote the sequence of continuous speech representations, and let $\mathbf{c}$ denote the text token conditions.
We model the joint distribution in an autoregressive manner as 
\begin{align}
    p(\mathbf{z}\mid \mathbf{c}) = \prod_{i=1}^{T} p(\mathbf{z}_i \mid \mathbf{c},  \mathbf{z}_{<i})
\end{align}
To mitigate the challenges of long-sequence modeling, we adopt a patch-based formulation by grouping consecutive frames into patches of size $L$.
Each patch $\mathbf{z}_{iL:(i+1)L-1}$ is treated as a single token $\mathbf{p}_i$ in the autoregressive process.

\textbf{LLM Backbone for Patch-wise Modeling}:
We feed the sequence of patch-level speech representations into a causal LLM to model long-range dependencies. 
The LLM processes the prefix $\mathbf{p}_{< i}$ and produces a hidden state $\mathbf{h}_{i-1} = \mathrm{LLM}(\mathbf{c}, \mathbf{p}_{<i})$, encoding the autoregressive context.

\textbf{Diffusion-based Next-Patch Generation}:
Instead of directly predicting the next patch $\mathbf{p}_i$, we model the conditional distribution of next continuous representation patch using a diffusion model, i.e., $\mathbf{p}_{i} \sim p_{\theta}(\cdot \mid \mathbf{h}_{i-1})$.
Following DiTAR~\cite{ditar}, we employ a lightweight Diffusion Transformer (DiT)~\cite{peebles2023scalable} as the diffusion head, referred to as Local Diffusion Transformer (LocDiT).
The bidirectional Transformer architecture of LocDiT provides full-receptive-field modeling within each patch.

To enhance local coherence of generation, the diffusion head is conditioned not only on the LLM hidden state $\mathbf{h}_{i-1}$, but also on historical patches. 
Specifically, previously generated patches $\mathbf{p}_{i-1}$ are concatenated with the noisy target patch and fed into LocDiT. 
This context-aware design improve generation quality by formulating the task as outpainting rather than independent patch generation.

\subsubsection{Diffusion Formulation and Training Objective}
We adopt the denoising diffusion probabilistic model (DDPM)~\cite{ho2020denoising} formulation over continuous representation patches.
The diffusion process operates on discrete timesteps $t \in \{1, \dots, T_{\text{diff}}\}$, where Gaussian noise is gradually added to each clean latent patch $\mathbf{p}_{i,0}$.
Given a clean latent patch $\mathbf{p}_{i,0}$, the forward diffusion process at timestep $t$ is defined as $\mathbf{p}_{i,t} = \sqrt{\bar{\alpha}_t}\,\mathbf{p}_{i,0} + \sqrt{1 - \bar{\alpha}_t}\,\epsilon$, where $\epsilon \sim \mathcal{N}(0, I)$ and $\bar{\alpha}_t = \prod_{s=1}^{t} (1 - \beta_s)$ is a cumulative product defined by a predefined noise schedule $\{\beta_s\}_{s=1}^{T_{\text{diff}}}$.
The diffusion head LocDiT $\epsilon_\theta(\mathbf{p}_{i,t}, t, \mathbf{h}_{i-1}, \mathbf{p}_{i-1})$ predicts the injected noise conditioned on the noisy latent, the discrete timestep embedding, and the autoregressive context from the LLM hidden state $\mathbf{h}_{i-1}$ as well as the previous patch $\mathbf{p}_{i-1}$.
The model is trained using the standard noise prediction objective:
\begin{equation}
\mathcal{L}_{\text{diff}} = 
\mathbb{E}_{\mathbf{p}_{i,0}, t, \epsilon}
\left[
\| \epsilon - \epsilon_\theta(\mathbf{p}_{i,t}, t, \mathbf{h}_{i-1}, \mathbf{p}_{i-1}) \|_2^2
\right].
\end{equation}
Meanwhile, the LLM that produces the conditioning hidden state $\mathbf{h}_i$ is jointly optimized to predict utterance-level termination, i.e., whether the current speech utterance should end, determining whether a termination token (e.g., $\langle E \rangle$) is emitted at the current step.

\subsubsection{LLM-Based Classifier-Free Guidance}
To improve conditional generation, we adopt classifier-free guidance (CFG)~\cite{ho2021classifier} with LLM-based conditioning in the diffusion head.
During training, the conditioning hidden state $\mathbf{h}_{i-1}$ is randomly replaced with a null embedding $\emptyset$ with a certain probability, enabling joint training of conditional and unconditional predictions.

At inference time, we combine the conditional and unconditional noise predictions as $\hat{\epsilon} = (1 + w)\,\epsilon_\theta(\mathbf{p}_{i,t}, t, \mathbf{h}_{i-1}, \mathbf{p}_{i-1}) - w\,\epsilon_\theta(\mathbf{p}_{i,t}, t, \emptyset, \mathbf{p}_{i-1})$, where $w$ is the guidance scale, which is set to 2.5 in all experiments. 
This LM guidance method improves conditional generation quality while requiring only a single forward pass through the LLM.

\section{Experiments Setup}
\subsection{Training Dataset}
We conduct experiments using a 150K-hour bilingual corpus composed of 100K hours from the open-source Emilia dataset~\cite{emilia} and 50K hours of internally collected speech data, containing 25K hours each of Chinese and English speech.
We report two SemaVoice configurations: one using Emilia only and another using the full 150K-hour corpus.
The VAE models are trained on a 20K-hour bilingual subset sampled from Emilia, with 10K hours each of Chinese and English speech.
For ablation studies, all experiments are conducted on the English subset of Emilia, which contains approximately 46.8K hours of speech (reported as 50K hours for simplicity), where the VAE is trained on a sampled 10K-hour subset. All models use 24 kHz audio.

\subsection{Experimental Settings}
\textbf{Model Setup}: The $\sigma$-VAE in SemaVoice adopts a mirror-symmetric encoder-decoder architecture.
The encoder comprises seven hierarchical stages of modified Transformer blocks, where self-attention is replaced by 1D depth-wise convolutions. 
Six downsampling stages compress 24 kHz waveforms into 15 Hz continuous latent representations, corresponding to a 1600× temporal reduction. 
The WavLM-large\footnote{https://huggingface.co/microsoft/wavlm-large}~\cite{chen2022wavlm} is employed for SFM guided alignment in VAE training.
Audio is resampled to 16 kHz before WavLM feature extraction.
The loss terms in $\mathcal{L}_{\mathrm{VAE}}$ follow~\cite{kumar2023dac}, while the alignment objective $\mathcal{L}_{\mathrm{align}}$ is adopted from~\cite{yao2025reconstruction}.
The downstream autoregressive TTS model is initialized from the pre-trained Qwen2.5-1.5B model\footnote{https://huggingface.co/Qwen/Qwen2.5-1.5B}~\cite{yang2024qwen2}.
We use a patch size of 2, and the previous patch is provided as conditioning context for the diffusion head during generation.

\textbf{Implementation Details}:
All VAE models are trained on 8 NVIDIA A800 GPUs for 280K steps with a global batch size of 320 seconds.
TTS models are trained on 8 NVIDIA H200 GPUs with a global batch size of 8192 seconds for 150K steps (SemaVoice-Emilia) or 300K steps (SemaVoice).
Ablation TTS models are trained on 8 NVIDIA A800 GPUs with a global batch size of 1024 seconds for 100K steps. 
We use cosine learning rate decay with a peak learning rate of $1\times10^{-4}$ for VAE and full-scale TTS training, and $7.5\times10^{-5}$ for all ablation TTS models. 

\subsection{Evaluation Metrics}
We evaluate SemaVoice using both objective and subjective metrics. For zero-shot TTS, objective metrics include Word / Character Error Rate (WER / CER) for intelligibility and robustness, and speaker similarity (SIM) for voice cloning performance. 
Subjective quality is measured by Mean Opinion Score (MOS), where 12 raters score naturalness (N-MOS) and speaker similarity (S-MOS) on a 5-point scale over 20 evaluation utterances.
All TTS evaluations are conducted on the public Seed-TTS benchmark~\cite{anastassiou2024seed}, which covers English and Chinese test sets, together with a more challenging Hard subset containing complex sentences. We use Whisper-large-v3\footnote{https://huggingface.co/openai/whisper-large-v3} for English ASR evaluation, and Paraformer-zh\footnote{https://huggingface.co/funasr/paraformer-zh} for Chinese and Hard subsets.
Speaker similarity is computed using a WavLM-large model\footnote{https://github.com/microsoft/UniSpeech/tree/main/WavLM}.
We further assess VAE reconstruction quality using STOI, PESQ, and UTMOS, measuring intelligibility, perceptual quality, and naturalness of reconstructed speech.

\subsection{Baselines}
We compare SemaVoice against a wide range of state-of-the-art open-source TTS systems, including F5-TTS~\cite{chen2025f5}, MaskGCT~\cite{wangmaskgct}, CosyVoice series~\cite{du2024cosyvoice,du2024cosyvoice2}, Spark-TTS~\cite{wang2025spark}, FireRedTTS series~\cite{guo2024fireredtts,xie2025fireredtts}, IndexTTS 2~\cite{zhou2026indextts2}, VoxCPM~\cite{zhou2026hierarchical}, and recent audio language models such as HiggsAudio-v2~\cite{higgsaudio2025} and Qwen2.5-Omni~\cite{xu2025qwen2}. Baseline results are obtained from official implementations or reported numbers under comparable evaluation settings.

\begin{table}[thbp]
\centering
\small
\setlength{\tabcolsep}{3pt} 
\caption{Objective evaluation results on the Seed-TTS testset. In the Type column, C/D denote continuous/discrete speech representations, and modeling paradigms of AR/NAR/MLLM denote auto-regressive/non-autoregressive/multi-modal LLM modeling respectively. $\uparrow$ and $\downarrow$ represent that higher or lower values are better.}
\label{tab:main_results}
\begin{tabular}{lccccccccc}
\toprule
\multirow{2}{*}{\textbf{Model}} & \multirow{2}{*}{\textbf{Type}} & \multirow{2}{*}{\textbf{Params}} & \multirow{2}{*}{\textbf{\#Hours}} & \multicolumn{2}{c}{\textbf{EN}} & \multicolumn{2}{c}{\textbf{ZH}} & \multicolumn{2}{c}{\textbf{Hard}} \\
\cmidrule(lr){5-6} \cmidrule(lr){7-8} \cmidrule(lr){9-10}
& & & & \textbf{WER} $\downarrow$ & \textbf{SIM} $\uparrow$ & \textbf{CER} $\downarrow$ & \textbf{SIM} $\uparrow$ & \textbf{CER} $\downarrow$ & \textbf{SIM} $\uparrow$ \\
\midrule

Ground Truth & - & - & - & 2.14 & 0.734 & 1.26 & 0.755 & - & - \\
\midrule
Qwen2.5-Omni & MLLM & 7.0B & - & 2.72 & 0.632 & 1.70 & 0.752 & 7.97 & \textbf{0.747} \\
\midrule

F5-TTS & C-NAR & 0.3B & 100K & 2.00 & 0.647 & 1.52 & 0.741 & 8.67 & 0.713 \\
MaskGCT & D-NAR & 1.0B & 100K & 2.62 & 0.717 & 2.27 & \textbf{0.774} & - & - \\
\midrule

SparkTTS & D-AR & 0.5B & 100K & 1.98 & 0.573 & 1.20 & 0.660 & - & - \\
FireRedTTS-2 & D-AR & - & 1.4M & 1.95 & 0.665 & 1.14 & 0.732 & - & - \\
OpenAudio-s1 & D-AR & 0.5B & 2.0M & 1.94 & 0.550 & 1.18 & 0.685 & 23.37 & 0.643 \\
HiggsAudio-v2 & D-AR & 3.0B & 10M & 2.44 & 0.677 & 1.50 & 0.740 & 55.07 & 0.656 \\
CosyVoice & D-AR+C-NAR & 0.3B & 170K & 4.29 & 0.609 & 3.63 & 0.723 & 11.75 & 0.709 \\
CosyVoice2 & D-AR+C-NAR & 0.5B & 170K & 2.57 & 0.659 & 1.45 & 0.757 & \textbf{6.83} & 0.724 \\
FireRedTTS & D-AR+C-NAR & 0.5B & 248K & 3.82 & 0.460 & 1.51 & 0.635 & 17.45 & 0.621 \\
IndexTTS 2 & D-AR+C-NAR & 1.5B & 55K & 2.23 & 0.706 & 1.03 & 0.765 & 7.12 & 0.755 \\
\midrule

VoxCPM-Emilia & C-AR & 0.5B & 100K & 2.34 & 0.681 & 1.11 & 0.740 & 12.46 & 0.698 \\
VoxCPM & C-AR & 0.5B & 1.8M & 1.85 & \textbf{0.729} & \textbf{0.93} & 0.772 & 8.87 & 0.730 \\
VibeVoice & C-AR & 1.5B & - & 3.04 & 0.689 & 1.16 & 0.744 & - & - \\
\midrule

\textbf{SemaVoice-Emilia} & C-AR & 1.5B & 100K & 1.91 & 0.657 & 1.32 & 0.728 & 9.37 & 0.687 \\
\textbf{SemaVoice} & C-AR & 1.5B & 150K & \textbf{1.71} & 0.694 & 1.18 & 0.754 & 8.09 & 0.711 \\

\bottomrule
\end{tabular}
\end{table}

\section{Experimental Results}

\subsection{Objective Evaluation of Zero-shot TTS}
Table~\ref{tab:main_results} presents an objective performance comparison on Seed-TTS eval benchmark between proposed SemaVoice models and baseline TTS systems.
Several trends can be observed:

1) SemaVoice achieves strong performance on the Seed-TTS evaluation benchmark and consistently outperforms a wide range of open-source TTS systems across different modeling paradigms. In terms of speech intelligibility, it obtains the lowest English WER of 1.71\% and a competitive Chinese CER of 1.18\%, demonstrating strong cross-lingual synthesis capability under zero-shot settings.

2) In terms of speaker similarity, SemaVoice achieves 0.694 in English and 0.754 in Chinese. Compared with discrete autoregressive systems, it demonstrates competitive performance across most baselines. Specifically, it is slightly below IndexTTS 2 in English (0.694 vs. 0.706), while in Chinese it remains closely matched to CosyVoice 2 (0.757) and IndexTTS 2 (0.765), indicating stable speaker consistency across languages.


3) On the challenging hard subset, SemaVoice achieves competitive performance (CER 8.09\% and speaker similarity 0.711), outperforming VoxCPM while remaining comparable to CosyVoice 2 and IndexTTS 2, demonstrating robust generation under difficult synthesis conditions.

4) Compared with large-scale audio foundation models such as Qwen2.5-Omni and HiggsAudio-v2, SemaVoice achieves consistently better speech quality on English and Chinese evaluation subsets, despite using a smaller model and data scale.

5) When compared with baseline models trained on similar data scales (e.g., F5-TTS, MaskGCT, VoxCPM-Emilia, and SparkTTS), SemaVoice-Emilia achieves strong speech intelligibility, obtaining the lowest English WER of 1.91\% among these systems, alongside a Chinese CER of 1.32\%. Meanwhile, it maintains comparable speaker similarity across both subsets.

\subsection{Subjective Evaluation of Zero-shot TTS}
\begin{table}[htbp]
\centering
\caption{Subjective Evaluations in terms of Naturalness (N-MOS) and Speaker Similarity (S-MOS). Results are reported with 95\% confidence intervals. We select the competitive baseline models for subjective comparison based on objective results.}
\label{tab:subjective_mos}
\begin{tabular}{lcccc}
\toprule
\multirow{2}{*}{\textbf{System}} & \multicolumn{2}{c}{\textbf{EN}} & \multicolumn{2}{c}{\textbf{ZH}} \\
\cmidrule(lr){2-3} \cmidrule(lr){4-5}
& \textbf{N-MOS} & \textbf{S-MOS} & \textbf{N-MOS} & \textbf{S-MOS} \\
\midrule
Ground Truth & 4.02 $\pm$ 0.09 & 4.53 $\pm$ 0.12 & 3.94 $\pm$ 0.10 & 4.45 $\pm$ 0.07 \\
CosyVoice 2 & 3.96 $\pm$ 0.13 & 3.78 $\pm$ 0.12 & 3.73 $\pm$ 0.11 & 4.01 $\pm$ 0.15 \\
IndexTTS 2 & 3.75 $\pm$ 0.11 & 3.93 $\pm$ 0.14 & 3.79 $\pm$ 0.13 & 4.07 $\pm$ 0.13 \\
\midrule
\textbf{SemaVoice-Emilia} & 3.86 $\pm$ 0.11 & 3.69 $\pm$ 0.12 & 3.91 $\pm$ 0.12 & 3.92 $\pm$ 0.12 \\
\textbf{SemaVoice} & 3.98 $\pm$ 0.12 & 3.89 $\pm$ 0.14 & 4.07 $\pm$ 0.13 & 4.03 $\pm$ 0.11\\
\bottomrule
\end{tabular}
\end{table}
Subjective evaluations (Table~\ref{tab:subjective_mos}) further support the objective results. 
SemaVoice achieves consistently strong performance across both languages in terms of naturalness and speaker similarity. 
On the English subset, SemaVoice attains an N-MOS of 3.98 $\pm$ 0.12, which is close to ground-truth quality (4.02 $\pm$ 0.09), and shows competitive performance compared to strong baselines such as IndexTTS 2 (3.75 $\pm$ 0.13), while maintaining speaker similarity comparable to strong baselines (CosyVoice 2 and IndexTTS 2).
On the Chinese subset, SemaVoice achieves the highest naturalness score of 4.07 $\pm$ 0.13, slightly surpassing all baseline systems, and reaches a competitive S-MOS of 4.03 $\pm$ 0.11, comparable to CosyVoice 2 (4.01 $\pm$ 0.15) and IndexTTS 2 (4.07 $\pm$ 0.13). 
These results suggest that SemaVoice provides a favorable balance between naturalness and speaker similarity across languages.

In comparison, SemaVoice-Emilia, trained on a smaller data scale, demonstrates competitive naturalness (3.86 $\pm$ 0.11 EN / 3.91 $\pm$ 0.12 ZH) but shows a relatively larger gap in speaker similarity, particularly on the Chinese subset. 
This indicates that while the proposed framework remains effective under limited data conditions, scaling up training data further improves speaker consistency and perceptual quality.

\subsection{Ablation Study on Key Design Components}




\begin{table}[ht]
\centering
\caption{Ablation study of key components in SemaVoice on the Seed-TTS English test set. History refers to conditioning the LocDiT module on the previously generated representation patch. All experiments are conducted on the Emilia-EN subset.}
\label{tab:core_ablation}
\begin{tabular}{lcccc}
\toprule
\textbf{Model Configuration} & \textbf{SFM Align.} & \textbf{History} & \textbf{WER$\downarrow$} & \textbf{SIM$\uparrow$} \\
\midrule
\textbf{SemaVoice} & \checkmark & \checkmark & \textbf{2.97} & \textbf{0.635} \\
\quad w/o SFM Align. & \texttimes & \checkmark & 3.40 & 0.625 \\
\quad w/o History & \checkmark & \texttimes & 8.46 & 0.587 \\
\bottomrule
\end{tabular}
\end{table}

Table~\ref{tab:core_ablation} presents an ablation study on the key components of SemaVoice, including the SFM-guided alignment and the historical conditioning in the LocDiT module.

Removing the SFM-guided alignment leads to a clear degradation in both intelligibility and speaker similarity, with WER increasing from 2.97\% to 3.40\% and SIM dropping from 0.635 to 0.625. 
This result demonstrates that semantic alignment between continuous speech representations and SFM features plays a critical role in improving both metrics. 
The consistent degradation across intelligibility and speaker similarity indicates that the proposed alignment mechanism effectively enhances the quality of the learned representation space.

In contrast, removing the historical conditioning results in a substantial performance degradation, with WER sharply increasing to 8.46\% and SIM decreasing to 0.587. 
This highlights the critical role of patch-level dependency modeling in the diffusion process. 
Without conditioning on the previously generated representation patch, the model loses local temporal and acoustic continuity, leading to inconsistent generation and accumulated errors across segments.

\subsection{Impact of Representation Granularity on Alignment Effectiveness}
\begin{table}[ht]
\centering
\caption{Effect of representation granularity under a fixed information rate on SFM-guided alignment. 
By varying the frame rate and latent dimensionality of the VAE while keeping the overall representation capacity approximately constant, we analyze how alignment behaves under different sequence modeling difficulties. 
Results are reported for reconstruction (LibriTTS test-clean) and zero-shot generation (Seed-TTS English test set).}
\label{tab:freq_ablation}

\begin{tabular}{c c c ccc cc}
\toprule
\multirow{2}{*}{\textbf{Freq.}} 
& \multirow{2}{*}{\shortstack{\textbf{Latent} \\ \textbf{Dim.}}}
& \multirow{2}{*}{\shortstack{\textbf{SFM} \\ \textbf{Align.}}} 
& \multicolumn{3}{c}{\textbf{Reconstruction (LibriTTS)}} 
& \multicolumn{2}{c}{\textbf{Generation (Seed-TTS)}} \\
\cmidrule(lr){4-6} \cmidrule(lr){7-8}
& & & \textbf{PESQ$\uparrow$} & \textbf{STOI$\uparrow$} & \textbf{UTMOS$\uparrow$} 
& \textbf{WER$\downarrow$} & \textbf{SIM$\uparrow$} \\
\midrule

\multirow{2}{*}{15Hz} 
& \multirow{2}{*}{32} 
& \checkmark & 3.175 & 0.950 & 4.030 & 2.97 & 0.635 \\
& 
& \texttimes & 3.179 & 0.949 & 4.066 & 3.40 & 0.625 \\
\midrule

\multirow{2}{*}{30Hz} 
& \multirow{2}{*}{16} 
& \checkmark & 3.086 & 0.953 & 4.052 & 3.24 & 0.621 \\
& 
& \texttimes & 3.078 & 0.953 & 4.060 & 5.20 & 0.615 \\
\midrule

\multirow{2}{*}{60Hz} 
& \multirow{2}{*}{8} 
& \checkmark & 2.908 & 0.945 & 3.965 & 14.71 & 0.526 \\
& 
& \texttimes & 2.880 & 0.944 & 3.965 & 28.06 & 0.464 \\
\bottomrule
\end{tabular}
\end{table}

Table~\ref{tab:freq_ablation} analyzes the effect of representation granularity on SFM-guided alignment under a fixed information rate. 
By increasing the frame rate while reducing the latent dimensionality, the representation becomes more fine-grained, resulting in longer sequences and increased difficulty for autoregressive modeling.
We first observe that reconstruction quality (PESQ, STOI, UTMOS) remains relatively stable across different configurations, indicating that the overall representation capacity is comparable under a fixed information rate. 
In contrast, as the representation becomes more fine-grained (from 15Hz to 60Hz), generation performance degrades significantly, reflecting the increased difficulty of autoregressive sequence modeling under higher temporal resolution. 
More importantly, the performance gap between models with and without SFM-guided alignment consistently enlarges as the representation becomes finer-grained. 
For example, at 60Hz, removing alignment leads to a substantial increase in WER from 14.71\% to 28.06\%, compared to a much smaller gap at 15Hz (2.97\% vs. 3.40\%).
This trend suggests that SFM-guided alignment becomes increasingly important as the representation granularity increases. 
When the sequence modeling burden is higher, the semantic guidance provided by the SFM helps stabilize the representation and improve generation robustness. 
In contrast, without alignment, the model struggles to maintain coherent semantic structure under more demanding modeling conditions.

\section{Conclusion}
In this work, we proposed SemaVoice, a semantic-aware continuous autoregressive framework for high-fidelity zero-shot text-to-speech synthesis.
At the core of the framework is an SFM-guided alignment mechanism, which addresses the mismatch between reconstruction-driven speech representations and semantic-prosodic modeling by explicitly aligning continuous latent representations with high-level semantic structures.
Extensive experiments demonstrate that SemaVoice achieves strong performance across both objective and subjective evaluations.
Further analysis shows that the proposed alignment mechanism becomes increasingly beneficial as the representation granularity increases, highlighting its role in stabilizing autoregressive generation under more demanding sequence modeling conditions.

\section{Limitation}
The evaluation is limited to bilingual datasets; extending it to more diverse languages and domains would further strengthen the generality of the approach.
As a continuous autoregressive framework, SemaVoice also faces inherent challenges such as sequential inference latency and potential error accumulation across long sequences.

\bibliographystyle{unsrtnat}
\bibliography{refs}

\end{document}